# Impact of helium ion irradiation on the thermal properties of superconducting nanowire single-photon detectors

**Yi-Yu Hong[1, 2, 3], Yu-Ze Wang[1, 2, 3], Wei-Jun Zhang[1, 2, 3*], Jia-Hao Hu[1, 2, 3], Jia-Min Xiong[1, 2, 3], Dong-Wei Chu[1, 2, 3], Xin Ou[1, 2, 3], Wen-Tao Wu[1, 2, 3], Xiao-Fu Zhang[1, 2, 3], Hui-Qin Yu[1, 2, 3], Pu-Sheng Yuan[1, 2, 3], Hao Li[1, 2, 3], Ling Wu[1, 2, 3], Zhen Wang[1, 2, 3], and Li-Xing You[1, 2, 3]**

[1]Shanghai Key Laboratory of Superconductor Integrated Circuit Technologies, 865 Changning Road, Shanghai 200050, China.

[2]National Key Laboratory of Materials for Integrated Circuits, Shanghai Institute of Microsystem and Information Technology, Chinese Academy of Sciences, 865 Changning Road, Shanghai 200050, China.

[3]Center of Materials Science and Optoelectronics Engineering, University of Chinese Academy of Sciences, Beijing 100049, China.

[*]Contact author: zhangweijun@mail.sim.ac.cn

**Abstract:** Superconducting nanowire single-photon detectors (SNSPDs) offer high system detection efficiency (SDE), low dark counts, and excellent timing resolution, making them indispensable for applications from quantum optics to deep-space optical communications. Although near-unity SDE and intrinsic detection efficiency (IDE) have been achieved at near-infrared wavelengths through advanced fabrication and optical design, further improving IDE at the material and device levels remains essential for achieving broader bandwidths, higher operating temperatures, and simplified processing.

Here, we explored helium ion irradiation as a post-fabrication method to modify the thermal properties of NbN SNSPDs without altering their geometry or material stack. With increasing fluence (0–1.1 × $10^{17}$ ions/cm$^2$), the thermal boundary conductance decreases by 57% from 127 W/m$^2$K$^4$ to 54 W/m$^2$K$^4$, and saturates near 9 × $10^{16}$ ions/cm$^2$. At this fluence, the minimum hotspot relaxation time increases by 41%, rising from 17 ps to 24 ps, while the electron–phonon interaction time decreases, with the magnitude of change depending on temperature and sample batch. Transmission electron microscopy reveals interfacial disorder near the NbN/$SiO_2$ interface (6–8 nm) and localized defects in the $SiO_2$ layer (30–260 nm), correlating with thermal transport suppression. These thermal modifications have a significant impact on relaxation dynamics and contribute to IDE enhancement. Our findings establish ion irradiation as an effective strategy for thermal tuning in SNSPDs, providing new insights into defect engineering for superconducting optoelectronic devices.

## 1. Introduction

Superconducting nanowire single-photon detectors (SNSPDs) are crucial for applications in quantum optics, quantum information, and deep-space optical communications, owing to their high system detection efficiency (SDE), low dark counts, and minimal timing jitter [1-4]. Compared to traditional single-photon avalanche diodes (SPADs) [5], SNSPDs offer superior performance in the near- and mid-infrared (MIR) ranges, enabling emerging applications in on-chip quantum systems [6], biomedical imaging [7], and dark matter detection [8].

Typical SNSPDs consist of 5–8 nm-thick NbN-films patterned into ~100 nm-wide nanowires and cooled to cryogenic temperatures (2–4 K). Recent efforts aim to expand their capabilities toward longer wavelengths (e.g., MIR [9, 10]), wider strips (e.g., microstrip devices [11, 12]), higher operating temperatures [13-15], and large-scale arrays [16-18]. While near-unity SDE and IDE have been achieved via nanowire geometry optimization and multilayer optical cavity designs [19-21], an alternative direction is to employ material- and process-level post-fabrication tuning strategies that can enhance device performance without introducing additional geometric or fabrication complexity.

Generally, IDE is governed by factors including the superconducting energy gap ($\Delta_0$), nanowire geometry, photon energy, and operating temperature [17–20]. Its dependence on bias current typically follows a sigmoid curve, with the saturated plateau indicating near-unity IDE. Common strategies to enhance IDE fall into four categories: (1) geometry scaling (thinner/narrower wires) [22], which lowers detection thresholds but also significantly reduces current critical temperature ($T_c$) or critical current ($I_c$); (2) material optimization, where amorphous materials such as $WSi_x$ [10] can enable high IDE but generally require lower operating temperatures due to their reduced $T_c$, while polycrystalline materials such as $NbTiN_x$ [15, 23] have demonstrated high currents and good performance at elevated temperatures, though their optimization often requires complex composition and growth control; (3) uniformity enhancement through advanced fabrication techniques [24, 25], which typically involve increased processing complexity; and (4) ion irradiation, a promising post-fabrication approach that enables precise defect engineering without modifying device geometry or optical stack.

In 2019, Zhang et al [26] showed that helium ion ($He^+$) irradiation can effectively enhance IDE of NbN SNSPDs by modifying key electrical parameters, including $\Delta_0$, the electronic density of states at the Fermi level ($N_0$), and the sheet resistance ($R_s$). Recently, this method has been extended not only to other superconducting materials, including NbTiN [27], MoSi [28], and high-$T_c$ materials [13, 14, 29], but also to studies that have deepened the understanding of the underlying mechanisms [27-29]. However, its influence on thermal transport, a critical aspect of detection dynamics, remains largely unexplored.

In SNSPDs, photon absorption breaks Cooper pairs and generates quasiparticles, whose cascade locally suppresses superconductivity to form a resistive hotspot. [30, 31]. The hotspot formation and relaxation dynamics involve several thermal timescales [32]: electron-electron scattering time ($\tau_{e\text{-}e}$), electron-phonon interaction time ($\tau_{e\text{–}ph}$) [33, 34], phonon escape time to the substrate ($\tau_{esc}$) governed by thermal boundary conductance (TBC) [35], and hotspot relaxation time ($\tau_{hs}$) [36]. These processes jointly determine IDE, response time (pulse rise time after photon absorption, reflecting hotspot dynamics), timing jitter (statistical fluctuation of detection times), and the detector's ability to reset (recovery time for the nanowire to return to the superconducting state). Efficient heat dissipation (i.e., short $\tau_{esc}$ and high TBC) allows faster operation, while suppressed phonon escape (i.e., low TBC) can improve energy confinement and IDE [37].

Thermal coupling is also influenced by substrate choice, with materials like $SiO_2$ and $Al_2O_3$ exhibiting different TBC [35] and $\tau_{hs}$ [38]. Previous attempts to modify thermal properties via nanowire suspension [39], mesa structures[40], or substrate variation [35, 38]—inherently alter device geometry or thin-film quality, complicating direct comparisons. In contrast, ion irradiation allows systematic before-and-after studies on the same device without structural changes. Nevertheless, key open questions remain: where do irradiation-induced changes occur—within the nanowire, at the interface, or in the substrate? How do they affect $\tau_{e\text{–}ph}$, $\tau_{esc}$ (TBC), and $\tau_{hs}$? Ultimately, how do these thermal changes impact detection performance?

In this work, we systematically investigate the effect of $He^{+}$ irradiation on the thermal properties of NbN-based SNSPDs. Using multiple experimental techniques, we characterized $\tau_{e\text{-}ph}$, TBC, and $\tau_{hs}$, and demonstrated how irradiation fluence modifies thermal relaxation pathways. Our TEM and SRIM analyses further reveal microstructural changes in the irradiated devices. Our findings establish a promising strategy for post-fabrication thermal tuning and deepen the understanding of SNSPD detection mechanisms, supporting the development of next-generation devices such as high-temperature and MIR-sensitive SNSPDs.

## 2. Methods

### 2.1. Sample preparation and ion irradiation simulation

Polycrystalline NbN thin films (6.5 or 7 nm thick) were sputter-deposited onto distributed Bragg reflector (DBR) substrates [21]. The DBR, consisting of 13 alternating $SiO_2/Ta_2O_5$ pairs (quarter-wave layers of ≈267 nm $SiO_2$ and ≈185 nm $Ta_2O_5$ at 1550 nm), enhances light absorption in the nanowires. Both meandering nanowire structures (devices) and nanowire bridges (nanobridges) were patterned on the same wafer. Scanning electron microscopy (SEM) images (Figs. 1a-1b) confirm wire widths of ~93–95 nm. The devices were designed with filling factors of 30% and 70%, corresponding to measured values of 31% and 73%, respectively, with an active area diameter of 18 μm. The nanobridges are ~11 μm in length.

The screened devices were mounted on an 8-inch wafer and irradiated with 30 keV $He^+$ ions using a 300 mm medium-current ion implanter (Nissin Co., Exceed 2300RD). The irradiation fluence ranged from $2\times10^{16}$ to $1.1\times10^{17}$ ions/cm², achieved via multiple implantation steps. For example, Device A1 underwent three sequential irradiations with fluences of $2\times10^{16}$, $7\times10^{16}$, and $2\times10^{16}$ ions/cm², respectively, yielding a cumulative fluence of $1.1\times10^{17}$ ions/cm². After each implantation, the physical properties were re-measured for comparative analysis.

In this study, "fluence" always refers to the total accumulated fluence at the time of measurement. The multi-step irradiation approach was adopted to provide experimental flexibility and reduce the risk of device degradation at high doses. This stepwise method also enabled systematic monitoring of material and device parameters as a function of cumulative fluence. Control experiments indicate that devices subjected to multi-step irradiation display thermal and electrical characteristics similar to those irradiated once at the same total fluence. $He^+$ ions were selected to minimize chemical reactions, while the 30 keV energy was chosen to avoid ion-induced etching. Notably, the IDE of the SNSPD increased from 82% to 99% at 1550 nm after irradiation with a fluence of $9\times10^{16}$ ions/cm² (see Section A of Suppl. Data for details).

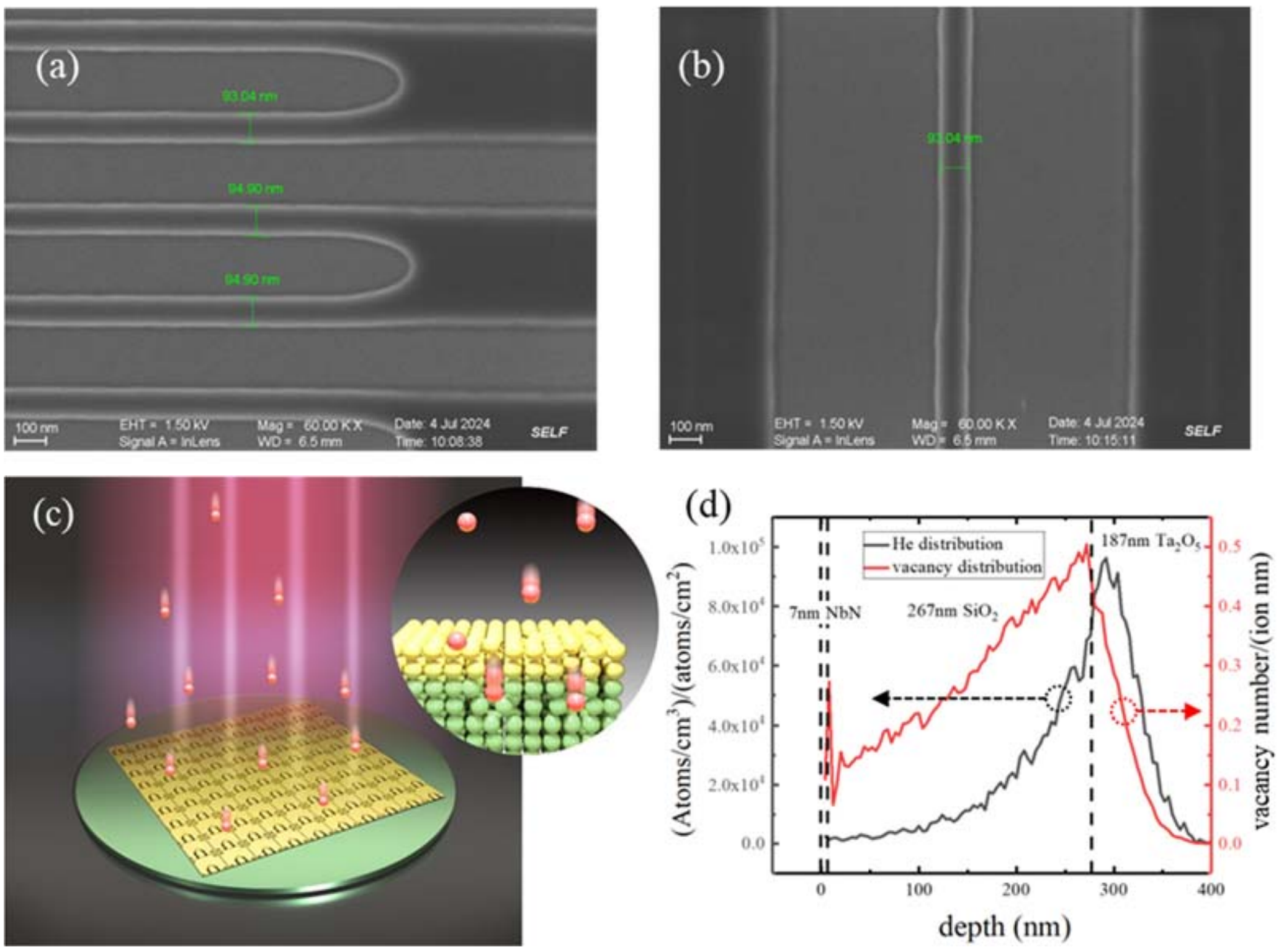

**FIG. 1.** SEM images of the samples: (a) meandered nanowires with circular bends; (b) a nanowire bridge; (c) Schematic illustration of an SNSPD wafer undergoing $He^+$ irradiation. The inset shows ions penetrating the material and generating vacancy defects; (d) Simulated depth profiles of $He^+$ concentration (black line) and vacancy defects (red dashed line) induced by ion irradiation.

Figure 1(c) is a schematic diagram of ion irradiation. Figure 1(d) displays the simulated depth profiles of implanted $He^+$ ions and associated vacancy defects, obtained using SRIM (Stopping and Range of Ions in Matter) calculations (see Section B of Suppl. Data). The $He^+$ ions fully penetrate the NbN thin film and predominantly accumulate near the interface of the first $SiO_2$/$Ta_2O_5$ bilayer, at a depth of ~ 272 nm. The vacancy distribution shows two distinct peaks: one near the NbN/substrate interface (~9 nm) and another within the $SiO_2$ layer (~292 nm). These results indicate the regions where irradiation-induced structural modifications are most likely to occur.

## 2.2. Analysis of thermal relaxation in SNSPDs

In SNSPDs, photon absorption initiates a thermal relaxation process that governs both the response time and detection efficiency. As shown in Figure 2, upon photon absorption, a Cooper pair is broken and high-energy quasiparticles are generated. Rapid electron–electron (e–e) scattering ((with a characteristic timescale $\tau_{e\text{-}e}$) redistributes the excess photon energy among electrons, establishing a locally thermalized population with an elevated effective temperature ($T_e$) and setting the initial hotspot size and formation speed. The hot-electron subsystem then transfers energy to the lattice via electron–phonon (e–ph) interactions, characterized by the relaxation time $\tau_{e\text{-}ph}$, which raises the phonon temperature ($T_{ph}$) and results in the formation of a resistive hotspot. Finally, the phonons dissipate their energy into the substrate through the thermal boundary conductance (TBC, denoted $\beta$). The overall thermal relaxation time ($\tau_{th}$) therefore reflects the combined effects of electron–phonon coupling and phonon escape processes within the two-temperature framework [40,43]:

$$\tau_{th} = \tau_{e-ph} + \left(1 + C_e/C_{ph}\right)\tau_{esc}, \tag{1}$$

where $C_e$ and $C_{ph}$ are the electron and phonon specific heats, and $\tau_{esc}$ is affected by the acoustic mismatch at the film–substrate interface. A smaller $\beta$ leads to a longer $\tau_{esc}$, thus increasing $\tau_{th}$.

It is important to distinguish between $\tau_{th}$ and $\tau_{hs}$. While $\tau_{th}$ reflects intrinsic thermal relaxation through electron–phonon coupling and phonon escape, $\tau_{hs}$ describes the recovery of superconductivity in the hotspot. This involves not only thermal processes but also electrothermal feedback [37] such as current redistribution, local Joule heating, and phonon escape. These combined effects make $\tau_{hs}$ a sensitive probe of both energy dissipation and superconducting recovery dynamics in SNSPDs.

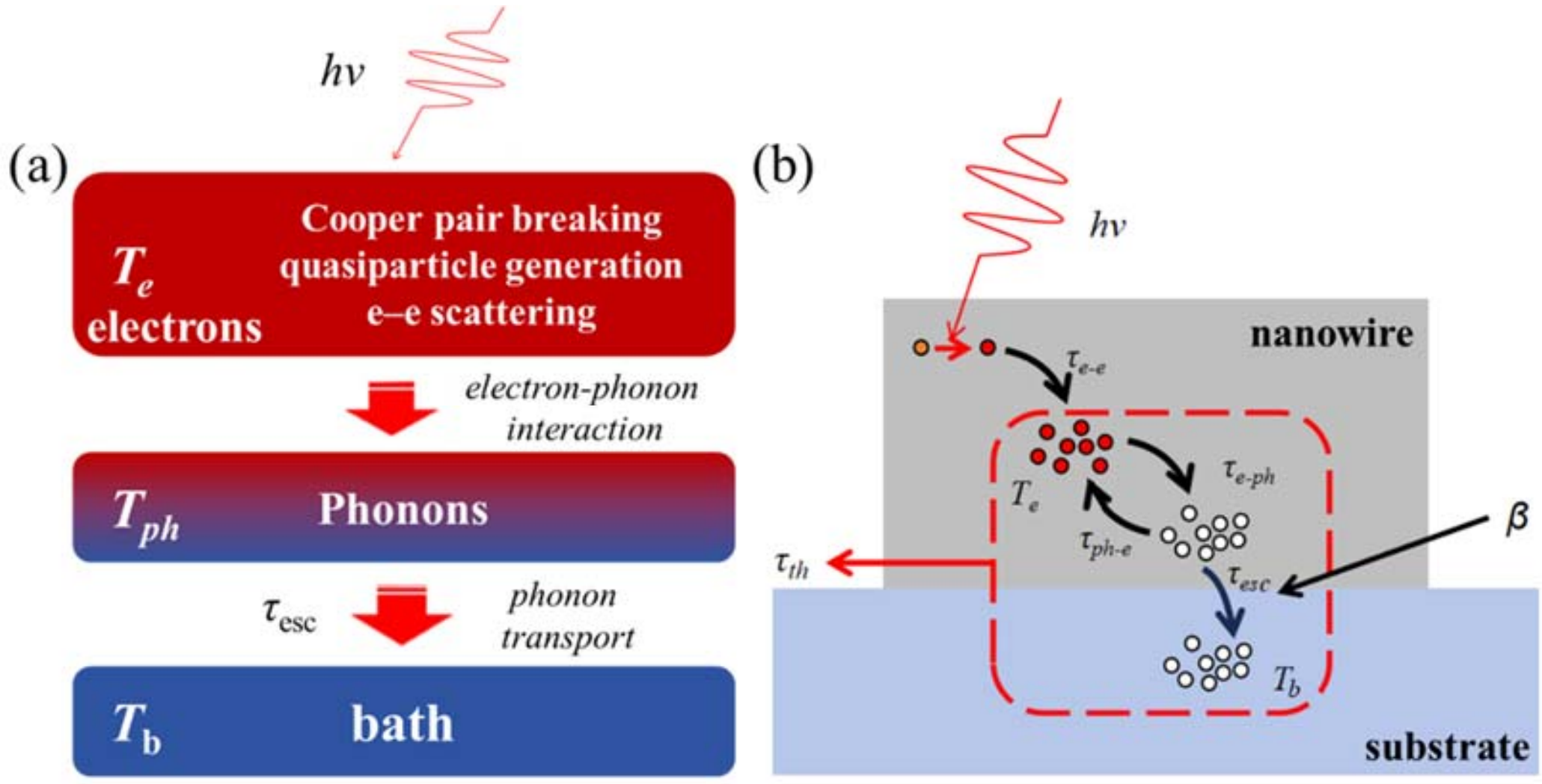

**FIG. 2.** Schematic of the thermal thermalization and relaxation process in the nanowire after photon absorption: (a) macroscopic view of photon energy conversion across different systems [32], where $T_e$ and $T_{ph}$ represent the effective temperatures of electrons and phonons, respectively, and $T_b$ is the bath temperature; (b) microscopic view highlighting key physical parameters [30], including e–e scattering and phonon interactions. $\beta$ refers to the TBC at the nanowire–substrate interface.

To experimentally probe these thermal processes, we employed three complementary techniques: magnetoconductance measurements to extract $\tau_{e\text{-}ph}$, self-heating hysteresis current analysis to estimate $\beta$, and two-photon pump–probe detection to determine $\tau_{hs}$. Each technique targets a distinct segment of the thermal pathway. By combining these methods, we construct a full-stack thermal picture of SNSPDs, enabling decoupled analysis of microscopic energy dissipation channels. The next sections present the corresponding measurements and comparative results before and after ion irradiation.

## 3. Results

### 3.1. Electron–phonon scattering time before and after ion irradiation

We evaluated $\tau_{e\text{-}ph}$ of NbN devices before and after $He^+$ irradiation using established magneto-transport methods [33, 41]. The analysis was based on excess magnetoconductance ($\delta\sigma$) measurements under perpendicular magnetic fields and involved fitting to standard localization and fluctuation conductivity models.

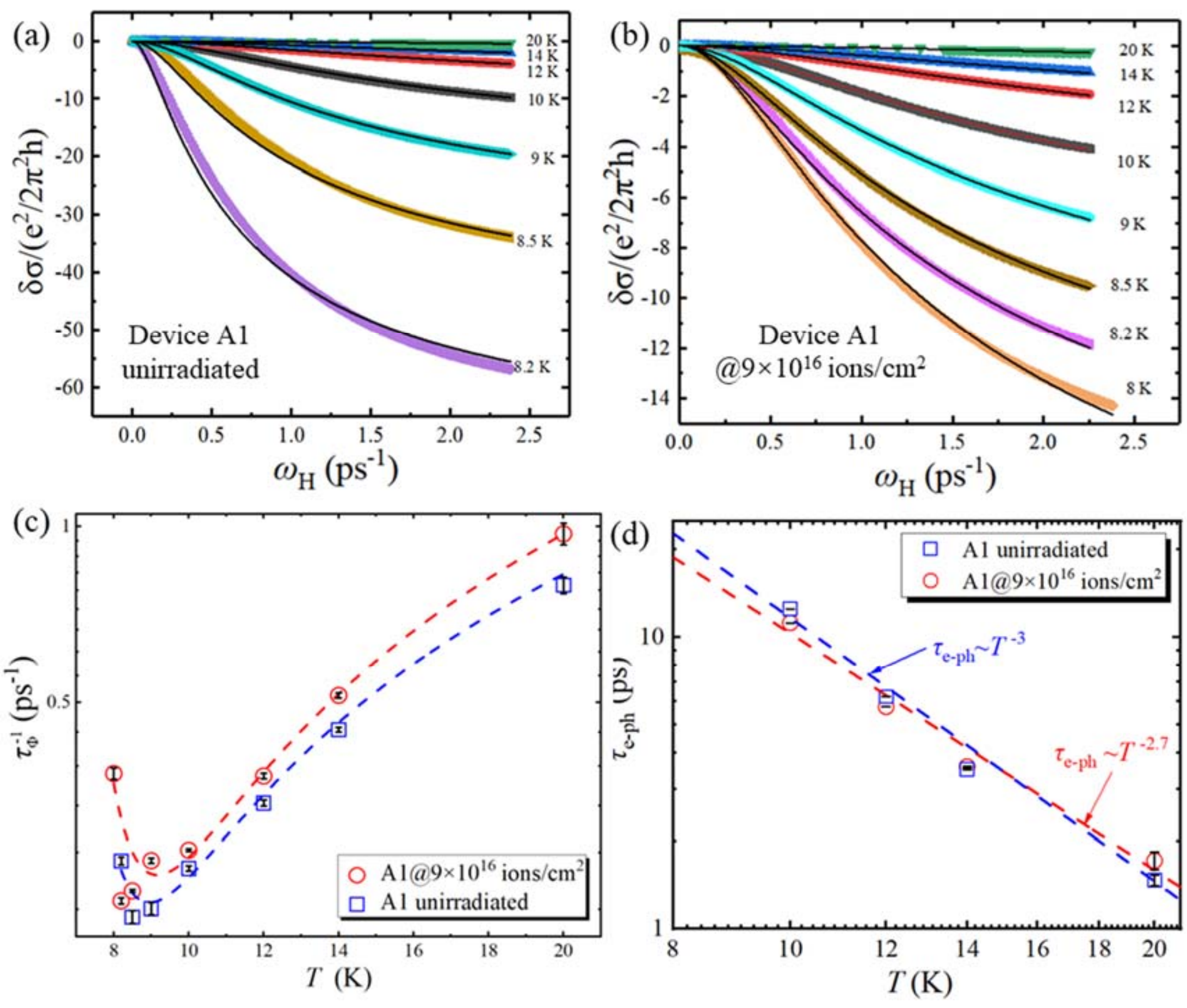

**FIG. 3.** Measurement data and-corresponding fits for the excess magnetoconductance of NbN films at various temperatures: (a)

before irradiation; (b) after irradiation; (c) temperature dependence of extracted $\tau_{\Phi}^{-1}$ for the devices before and after irradiation, where dashed lines are fitted results; (d) temperature dependence of extracted $\tau_{e\text{-}ph}$ for the devices before and after irradiation, plotted on a log-log scale. The dashed lines are fitted with $\tau_{e\text{-}ph} \sim T^{-n}$, where $n$ is approximately 3.0 for unirradiated and 2.7 for irradiated samples, respectively. Black error bars are shown in (c) and (d).

Figures 3(a) and 3(b) show the normalized excess magnetoconductance of Device A1 before and after irradiation at a fluence of $9 \times 10^{16}$ ions/cm² as a function of the cyclotron frequency $\omega_H = 4eD_eH/\hbar$ across different temperatures. $\omega_H$ is calculated in a disordered conductor with the reduced Planck's constant ($\hbar$) and the electron diffusion coefficient ($D_e$) of normal-state electrons. As $\omega_H$ increases, the magnitude of the negative excess magnetoconductance increases. The contributions to $\delta\sigma$ from weak localization, Maki-Thompson, Aslamazov-Larkin, and density of states effects are described in Ref. [33]. The experimental data fit well with the collective interplay of these four mechanisms. From the fits, we extracted the electron dephasing time $\tau_{\Phi}$ [34], and subsequently $\tau_{e\text{-}ph}$, as shown in Figure 3(c–d).

After irradiation, in Fig. 3(c), the average increase in $\tau_{\Phi}^{-1}$ is ~14% at 10–14 K, suggesting enhanced electron-phonon interaction due to irradiation-induced disorder. Correspondingly, in Fig. 3(d), $\tau_{e\text{-}ph}$ exhibits a power-law dependence $\propto T^{-n}$ (appearing nearly linear on the log–log scale), with the exponents slightly decreasing from ~ 3.0 to ~ 2.7 after irradiation. This fitting exponent $n$ aligns with previous studies on disordered NbN films. For instance, Sidorova et al. [41] reported $n \approx 3.2\text{–}3.8$ in highly disordered NbN films.

Here, $\tau_{e\text{-}ph}$ shows relatively small variations induced by irradiation. For instance, at 10 K, $\tau_{e\text{-}ph}$ decreases from 12.8 ps to 11.2 ps (~13%), while an independent sample set shows a larger reduction (~34% at 10 K, see Section I of the Suppl. Data). Despite these variations—likely attributable to intrinsic batch-to-batch differences such as film microstructure or defect density—$\tau_{e\text{-}ph}$ exhibits a consistent decreasing trend with increasing fluence (more evident in the Suppl. Data), indicating that $He^+$ irradiation reliably enhances electron–phonon coupling in NbN SNSPDs.

### 3.2. Thermal boundary conductance before and after ion irradiation

To assess the impact of irradiation on phonon escape processes, we evaluated TBC using a self-heating method based on the Skocpol-Beasley-Tinkham (SBT) framework [35, 40]. This framework assumes that phonon black-body radiation at the nanowire-substrate interface is the dominant hotspot cooling mechanism, with heat dissipation following a $T^4$ dependence. Thus, when the hotspot reaches thermal equilibrium, this leads to the modified expression:

$$\beta w^2(T_{hs}^4 - T_b^4) = I_{hs}^2 R_s \ (|x| < x_N), \tag{2}$$

where $\beta$ and $T_{hs}$ are fitting parameters; $w$ is the nanowire width; $T_{hs}$ is the hotspot temperature; $I_{hs}$ is the hotspot (or hysteresis) current, defined as the constant-current plateau at nonzero voltage during the downward sweep of the current–voltage (I–V) curve (see Section C of Suppl. Data).

Figure 4(a) shows magnified I–V curves measured at different temperatures, where slight fluctuations in the hysteresis current $I_{hs}$ appear more prominent due to the zoomed-in scale. Quantitatively, the peak-to-peak variation is typically 4–9% of the average hotspot current, while the standard deviation within the plateau is below ~2%. These fluctuations mainly arise from nanowire inhomogeneities (e.g., defects or width constrictions), which cause slight variations in the hotspot resistance and hence in the measured $I_{hs}$. To minimize their impact, $I_{hs}$ is determined by averaging over a region near zero bias within the hysteresis plateau (typically 300 data points; see Suppl. Data for more details), rather than relying on a single data point. The averaged value at 5.3 K is indicated as an example. Figure 4(b) displays the temperature dependence of the averaged $I_{hs}$, together with the corresponding fits, before and after irradiation. Both the switching current ($I_{sw}$) and $I_{hs}$ decrease after irradiation.

From the fits in Figure 4(c), the extracted TBC for unirradiated samples ranges from ~126 to 130 $W/m^2K^4$, while irradiated devices show reduced values between 51 and 63 $W/m^2K^4$. This variation likely arises from differences in thin-film quality and interface bonding strength, which affect phonon transport efficiency. Figure 4(d) summarizes this trend across multiple fluences, revealing saturation of TBC degradation beyond ~$9 \times 10^{16}$ ions/cm$^2$, likely due to defect saturation or limits in interfacial phonon transport. Statistically, the TBC decreases from ~127 $W/m^2K^4$ (unirradiated) to ~54 $W/m^2K^4$ at the highest fluence, corresponding to a ~57% reduction (or a factor of ~2.4). Notably, the TBC exhibits clear fluence dependence, suggesting that controlled ion irradiation offers a viable strategy for tuning thermal dissipation in SNSPDs.

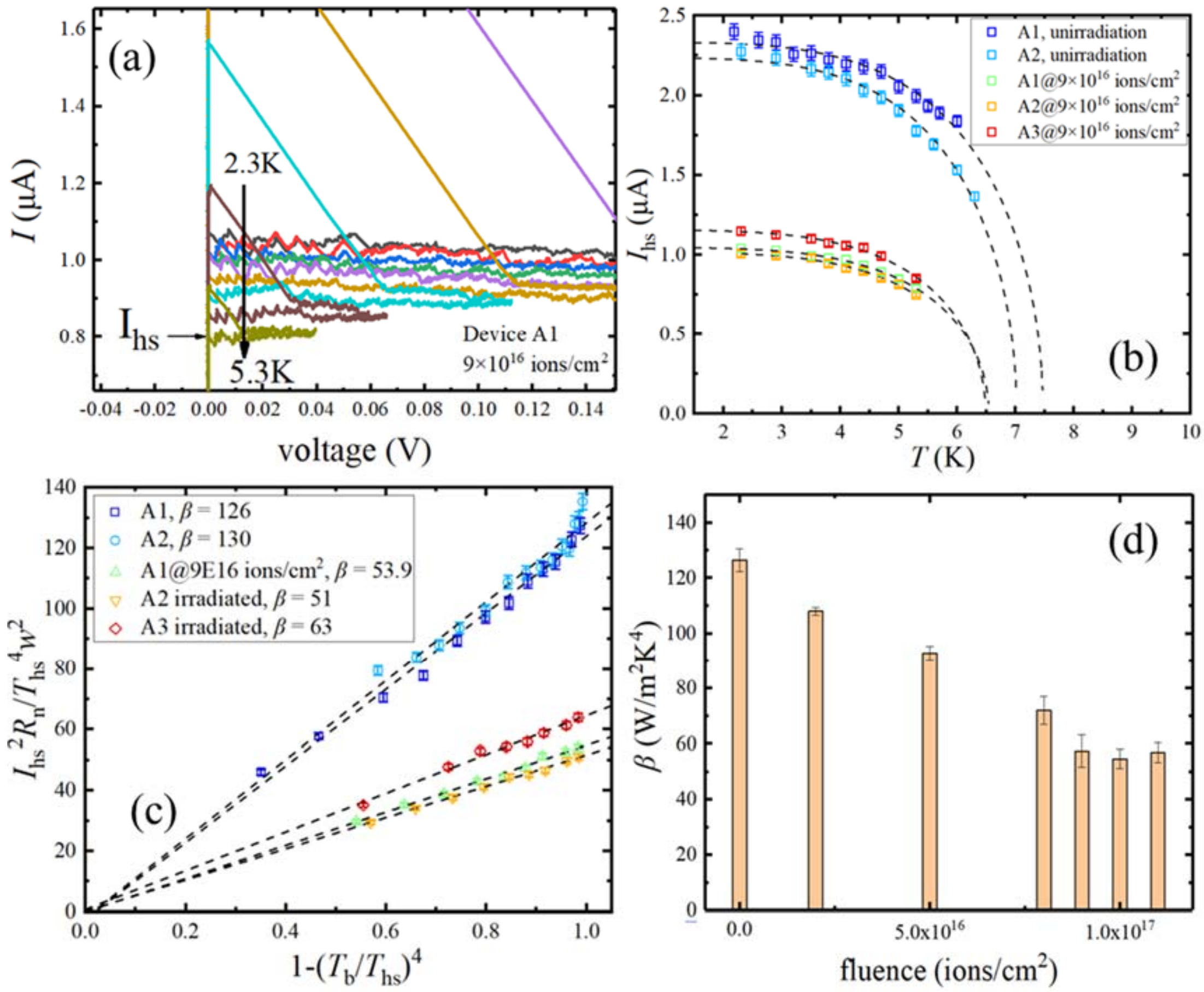

**FIG. 4.** (a) Magnified I-V characteristics of irradiated device A1 ($9 \times 10^{16}$ ions/cm²) measured at various temperatures. The hotspot current $I_{hs}$ at 5.3 K is indicated by the arrow. (b) Temperature dependence of $I_{hs}$ for devices before and after irradiation. Devices A1 to A3 were irradiated, while Devices D1 to D3 represent unirradiated devices. (c) Result of the coordinate axis transformation from Fig. 2(d), where the slope of the linear fit corresponds to the TBC (i.e., $\beta$) of the tested device. (d) Variation trend of the fitted TBC for the same batch of devices under different fluences, with average values (bars) and error bars (±5%).

### 3.3. Hotspot relaxation time before and after ion irradiation

We evaluated $\tau_{hs}$ using the two-photon pump–probe method [36], where the detection probability of time-delayed photon pairs reflects hotspot relaxation dynamics. Experimental details are provided in Section D of Suppl. Data. Figure 5(a) confirms two-photon detection at low bias currents (0.40 to 0.45 $I_{sw}$), where the photon count rate (PCR) slope is close to 2. Figure 5(b) shows representative normalized PCR as a function of pulse delay time ($t_d$) after irradiation. The half width at half maximum (HWHM) of the fitted Lorentzian curve defines $\tau_{hs}$.

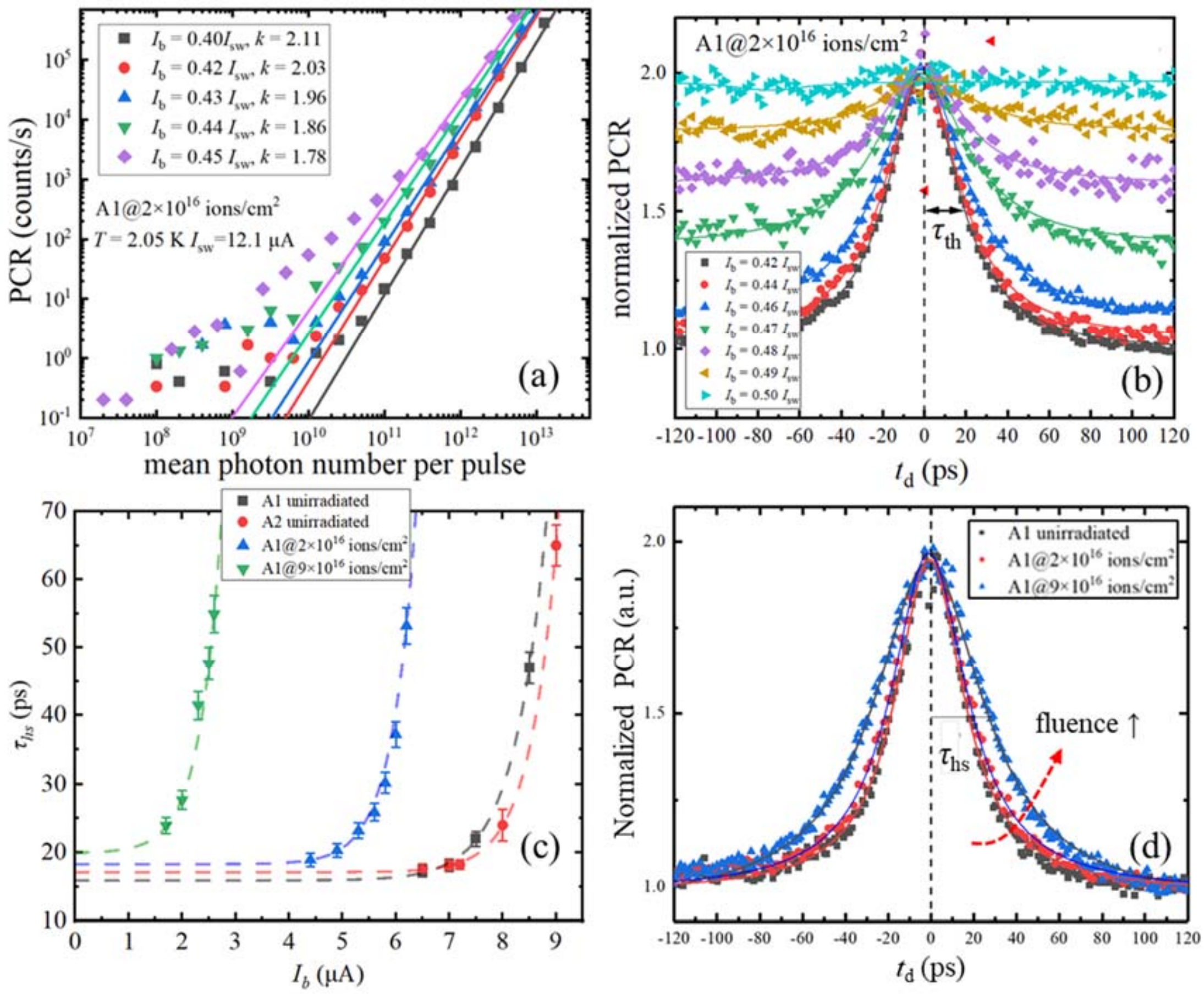

**FIG. 5.** (a) Photon count rate (PCR) as a function of mean photon number per pulse at different bias currents, with the fitted lines indicating the two-photon detection regime. (b) Normalized PCR as a function of delay time ($t_d$) after irradiation. (c) Comparison of hotspot relaxation time ($\tau_{hs}$) before and after irradiation across various devices as a function of bias current. Dashed lines are experiential decay function fits. (d) Normalized PCR as a function of $t_d$ for device A1 corresponding to $\tau_{hs\text{-}min}$ at each irradiation fluence. The dashed arrow indicates that the fluence increases.

Figure 5(c) displays $\tau_{hs}$ versus bias current for devices before and after irradiation. A pronounced increase in $\tau_{hs}$ is observed after irradiation, even at lower bias currents where the generated Joule heating is expected to be reduced. For instance, in unirradiated devices (A1 and A2), $\tau_{hs\text{-}min}$ was ~17.3 ps at 6.5 μA, with fitted values ranging from 15.9 to 16.9 ps. After irradiation with fluences of $2\times10^{16}$ and $9\times10^{16}$ ions/cm², $\tau_{hs\text{-}min}$ increased to 19.6 ps at 4.86 μA and 24 ps at 1.75 μA, respectively. Here, $\tau_{hs\text{-}min}$ is defined as the minimum hotspot relaxation time in each $\tau_{hs}$–$I_b$ curve, which corresponds to the value at the lowest measured bias since $\tau_{hs}$ generally increases with bias current. Figure 5(d) shows the measured and fitted Lorentzian curves used to extract $\tau_{hs\text{-}min}$ for Device A1 under each irradiation fluence. These results clearly demonstrate a clear fluence-dependent increase in $\tau_{hs}$, confirming that irradiation suppresses thermal relaxation in SNSPDs.

### 3.4. Analysis of microstructural changes in NbN thin films induced by irradiation

To investigate the effects of ion irradiation on the microstructure of the samples, we conducted transmission electron microscopy (TEM, JEM-2100F, JEOL) analyses on samples before and after irradiation. Figures 6(a) and 6(b) present the magnified TEM images near the NbN/$SiO_2$ interface and the corresponding fast Fourier transform (FFT) of each layer (left insets), respectively. Before irradiation, the interface between NbN and $SiO_2$ is visible, with an NbN thickness of 6.5–7.4 nm and a surface oxidation layer thickness of 0.7–2.0 nm. After irradiation, the interface between the two layers becomes blurred, accompanied by disorder. Figure (d) and its corresponding FFT analysis reveal the formation of a defect-rich layer in the $SiO_2$ near the NbN side, with a thickness of 5.8–8.0 nm. These observations are consistent with the SRIM simulation results presented in Section B of Suppl. Data, which show a primary subpeak in vacancy defect density located at the NbN/$SiO_2$ interface.

Moreover, FFT analysis reveals a mixture of polycrystalline and amorphous features in the NbN film before and after irradiation, without evidence of well-ordered crystalline domains. No significant grain restructuring or large-scale amorphization is observed, suggesting that irradiation causes only limited crystallographic damage in the NbN layer. However, irradiation-induced point defects or short-range disorder—beyond the resolution of conventional TEM—may still exist. This interpretation is supported by electrical measurements (Section G of Suppl. Data) and $\tau_{e\text{-}ph}$ analysis (Subsection 3.1), which show a ~1.6× increase in $R_s$ and an 13% decrease in $\tau_{e\text{-}ph}$, indicating moderately enhanced lattice scattering.

Figures 6(b), 6(c), 6(e), and 6(f) further illustrate the microstructural evolution across a broader depth range. For instance, the interface between $SiO_2$ and $Ta_2O_5$ is shown in Figure 6(f), where no significant vacancy defect layer is observed within a depth of approximately 450 nm. However, some dark contrast regions are visible, which are possibly related to local defect accumulation. Their density increases with irradiation fluences, and they are distributed in the range of 30–260 nm (Section E of Suppl. Data). Notably, the formation of He-induced bubbles in amorphous $SiO_2$ is generally suppressed due to its high helium diffusivity and absence of stable defect traps [42], in contrast to crystalline substrates such as Si [43].

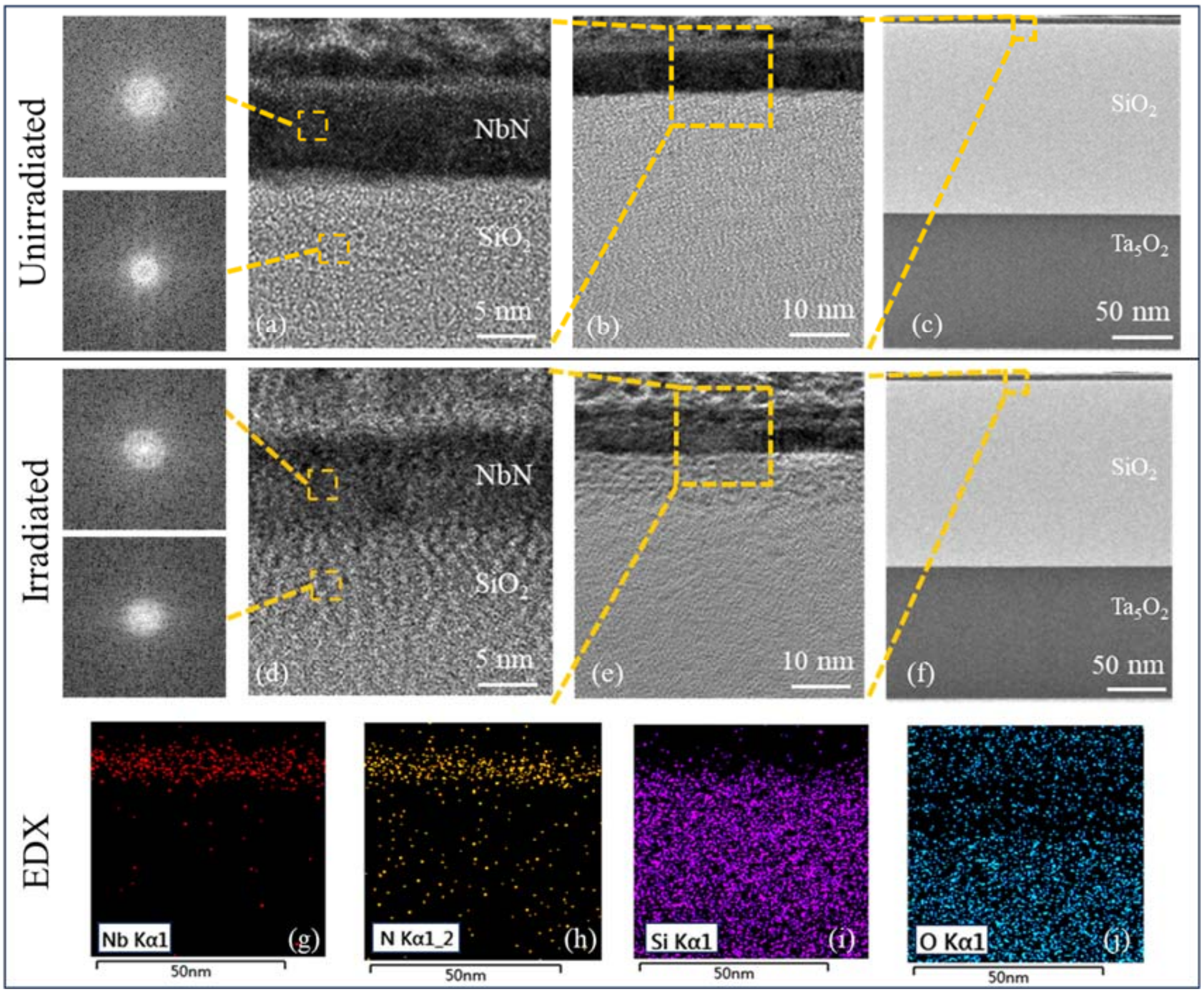

**FIG. 6.** TEM cross-sectional images of NbN thin films, comparing unirradiated and irradiated samples ($9\times10^{16}$ ions/cm²), taken from non-nanowire regions of the chips. (a) & (b), (c) & (d), and (e) & (f) correspond to scale bars of 5 nm, 10 nm, and 50 nm, respectively. A protective platinum layer was deposited atop the NbN film for TEM imaging. Insets of (a) and (b) are the FFT analysis for the NbN layer and $SiO_2$ layer, respectively. (g)-(j): EDX maps for Nb, N, Si, and O elements, respectively.

Figures 6(g)–6(j) present the energy-dispersive X-ray spectroscopy (EDX) mapping of the elemental composition in the irradiated films. The displaced Nb atoms are relatively sparse (Figure 6(g)) due to their high atomic mass, making them less susceptible to displacement under irradiation by low-mass He ions. In contrast, a substantial number of displaced N atoms are observed (Figure 6(h)), widely distributed within the $SiO_2$ layer to a depth of approximately 50 nm. The Si element is primarily concentrated in the $SiO_2$ layer (Figure 6(i)), while the O element exhibits a broader distribution, with a higher density of displacements or vacancies appearing at a depth of 4–14 nm beneath the NbN layer (Figure 6(j)). The EDX mapping results are consistent with the SRIM simulation (Section B of Suppl. Data).

After completing our initial experiments, we came across a recent study on arXiv [44] that reported TEM observations of NbTiN/$SiO_2$/Si multilayers subjected to $He^+$ ion microscope irradiation. Their findings indicated that high-fluence irradiation ($2 \times 10^{17}$ ions $cm^{-2}$) induced significant amorphization at the $SiO_2$/Si interface, extending to a depth of 350 nm, where a distinct deep bubble layer was observed. In contrast, our TEM results reveal that the most pronounced structural modifications occur at the NbN/$SiO_2$ interface. Additionally, we observed dark contrast regions within the $SiO_2$ layer; while these features may suggest localized irradiation-induced changes, their precise origin remains unclear and is unlikely to be classical He bubble formation due to the amorphous nature of $SiO_2$. We attribute the observed differences to the lower irradiation fluence used in our study and the difference in substrate crystallinity and layer stacking compared to Ref. [44].

## 4. Discussions

We summarize the impact of $He^+$ irradiation on the thermal transport properties of NbN devices and compare the results with previous reports (see Section F of Suppl. Data). For two independent sample batches, the pre-irradiation $\tau_{e\text{-}ph}$ at 10 K (12.8 ps and 13.8 ps) agrees with reported values for pristine NbN/$SiO_2$ [41], and decreases after $He^+$ irradiation to 11.2 ps (~13%) and 9.1 ps (~34%). These results confirm that irradiation enhances electron–phonon coupling, while the magnitude of change varies between batches. The TBC dropped by 57%, from 127 to 54 $W/m^2{\cdot}K^4$, in agreement with trends observed in irradiated NbTiN stacks at higher fluences [44]. $\tau_{hs\text{-}min}$ increased from 17 to 24 ps, marking a 41% rise, which also aligns with results from nanowire suspension structures [39].

While each parameter remains within the range reported for unmodified systems, to our knowledge, this is the first study to systematically track irradiation-induced thermal changes across multiple parameters in multilayer stacks. Notably, the 41% increase in $\tau_{hs\text{-}min}$ is comparable to changes achieved through more complex methods: e.g., substituting $SiO_2$ with Si raises $\tau_{hs\text{-}min}$ by ~95% [38], and nanowire suspension can double it [39]. In contrast, $He^+$ irradiation offers a post-fabrication means to tailor thermal transport, thereby providing a promising route for tuning the detection performance of SNSPDs. A schematic in Section H of the Suppl. Data illustrates how irradiation-induced disorder may hinder phonon transport. In polycrystalline NbN films, especially after ion irradiation, thermal bottlenecks result from the combined effects of acoustic mismatch at the NbN/$SiO_2$ interface and defect-enhanced phonon scattering on the ultrafast timescales relevant to hotspot relaxation.

In addition to thermal transport effects, we observe irradiation-induced modifications in electrical properties (see Table 1 and Suppl. Data), such as a ~30% reduction in $N_0$, a 58% increase in $R_s$, and a ~10% reduction in $\Delta_0$. The electron–electron interaction time ($\tau_{e\text{-}e}$) decreases by 10~20% after irradiation at $T_c$ for Device A1 and F1–F3 (Sections G and I, Supplementary Data), comparable in magnitude to the changes in electron–phonon interaction time. This suggests that modified e–e dynamics also contribute to the observed IDE enhancement [31].

For comparison, in Reference [43], the authors irradiated areas adjacent to the nanowire rather than the nanowire itself. As acknowledged in that study, a small fraction of He ions may still reach the nanowire via lateral or backscattering, so direct effects cannot be completely excluded.

Taken together, these observations indicate that IDE enhancement under irradiation arises from a complex interplay of reduced thermal conductance and modified electronic dynamics. While our study does not establish a strict one-to-one correspondence between thermal transport and detection efficiency, the combined tracking of thermal and electrical variations provides a consistent framework to interpret irradiation effects. Overall, $He^+$ irradiation serves as a versatile post-fabrication approach to simultaneously tune both channels in SNSPDs.

**Table 1.** Contributions of thermal and electrical parameters to the enhancement of IDE after irradiation ($9 \times 10^{16}$ ions cm$^{-2}$). Here, $N_0$ denotes the electronic density of states at the Fermi level, $D_e$ the electron diffusion coefficient, $R_s$ the sheet resistance of the devices, and $\Delta_0$ the superconducting energy gap at 0 K.

| **Performance category** | **Main parameter changes** | **Impact on IDE** |
|---|---|---|
| Thermal | TBC ↓57%, $\tau_{hs\text{-}min}$ ↑ 41%, $\tau_{e\text{-}ph}$ ↓ 13% | • Enhanced heat localization<br>• Increased quasiparticle lifetime |
| Electrical | $N_0$ ↓ 30%, $D_e$↓ 6%, $R_s$ ↑ 58%, $\Delta_0$ ↓ 10% | • Smaller $N_0$ and $D_e$ improve thermalization<br>• Reduced $\Delta_0$ and larger $R_s$ → hotspot growth |

Looking ahead, our findings suggest that both intrinsic material properties and thermal transport channels are critical for optimizing SNSPD sensitivity, with hotspot growth strongly promoted by higher sheet resistance and restricted heat dissipation. Integrating ion irradiation with interface or substrate engineering, such as suspended dielectric layers [39] or mesa structures [40], may further optimize thermal management in SNSPDs. Detailed modeling of phonon-mediated heat transport across interfaces can reveal TBC bottlenecks and guide device-level improvements. These insights provide a thermal-engineering perspective, alongside material composition control, as a promising strategy for developing high-$T_c$ and mid-infrared SNSPDs.

## 5. Conclusion

In summary, we demonstrate that $He^+$ irradiation is an effective post-fabrication strategy for tailoring the thermal properties of NbN thin films and enhancing the IDE of SNSPDs. By systematically characterizing key thermal parameters before and after irradiation, we observe a pronounced reduction in thermal boundary conductance, a moderate decrease in electron–phonon interaction time, and a significant extension of the hotspot relaxation time. TEM and SRIM analyses reveal the formation of a defect-rich NbN/$SiO_2$ interface and defect accumulation in the underlying $SiO_2$ layer, which together give rise to thermal bottlenecks through interfacial mismatch and defect-induced phonon scattering. These thermal modifications contribute to improved IDE performance. Our findings establish ion irradiation as a promising route for engineering SNSPDs with tunable thermal dynamics and improved sensitivity, while also providing a framework for defect-engineered superconducting optoelectronics.

**Acknowledgments:** This work is supported by the Innovation Program for Quantum Science and Technology (Grant No. 2023ZD0300100), the National Natural Science Foundation of China (Grant Nos. 62371443 and 61971409), and the Shanghai Sailing Program (Grant No. 22YF1456500). W.-J. Zhang is supported by an outstanding member of the Youth Innovation Promotion Association, CAS (Y2023071). The authors thank Ruo-Yan Ma, Xiao-Yu Liu, Bao Gao, and Min Zhou for their technical support.

**Author contributions:** W.-J.Z. conceived and designed the experiments, H.-Y.Y., Y.-Z.W., and W.-J.Z. fabricated the samples, conducted the experiments, and collected the data. H.-Q. Y. and W. -T. W. assisted with cryogenics measurements. H.-Y.Y. and W.-J.Z. analyzed the data and prepared the manuscript. All authors discussed the results and reviewed the manuscript.

**Competing interests:** The authors declare no competing interests.

**Data and materials availability:** All data that support the findings of this study are included within the article (and any supplementary files).